\documentstyle[prl,aps,epsf,floats]{revtex}
\begin{document}
\draft
\twocolumn[\hsize\textwidth\columnwidth\hsize\csname @twocolumnfalse\endcsname
\title{Short-range spin correlations and induced local spin-singlet amplitude \\
       in the Hubbard model}
\author{Bumsoo Kyung}
\address{D\'{e}partement de physique and Centre de recherche 
sur les propri\'{e}t\'{e}s \'{e}lectroniques \\
de mat\'{e}riaux avanc\'{e}s. 
Universit\'{e} de Sherbrooke, Sherbrooke, Qu\'{e}bec, Canada J1K 2R1}
\date{September 25, 2000}
\maketitle
\begin{abstract}

   In this paper, from the microscopic Hubbard Hamiltonian
we extract the local spin-singlet amplitude due to 
short-range spin correlations, and quantify its strength  
near half-filling.
As a first application of the present approach, we study a
problem of the energy dispersion and its d-wave modulation 
in the insulating cuprates, 
Sr$_{2}$CuO$_{2}$Cl$_{2}$ and 
Ca$_{2}$CuO$_{2}$Cl$_{2}$.
Without any adjustable parameters, 
most puzzling issues are 
naturally and quantitatively explained within the present approach.
\end{abstract}
\vskip2pc]
\narrowtext

    Recent discovery of
a normal state pseudogap in underdoped copper oxides\cite{Timusk:1999} 
has attracted
considerable attention both from 
experimentalists and theoretical physicists
for many years.
For these materials, the low frequency spectral weight begins to
be strongly suppressed
below some characteristic temperature
$T^{*}$ higher than $T_{c}$.
This anomalous behavior has been observed through various experimental
probes such as 
angle resolved photoemission spectroscopy
(ARPES),\cite{Ding:1996,Loeser:1996}
specific heat,\cite{Loram:1993}
tunneling,\cite{Renner:1998}
NMR,\cite{Takigawa:1991} and
optical conductivity.\cite{Homes:1993}
One of the most puzzling questions in pseudogap phenomena is why
$T^{*}$ has a completely different doping dependence from $T_{c}$,
in spite of possibly their close relation.

   There are many theoretical scenarios to attempt to understand the 
pseudogap phenomena.
These include 
the spinon pair formation without
the Bose-Einstein condensation of holons
in the slave boson (or fermion) mean-field theory,%
\cite{Anderson:1987,Baskaran:1987}
strong superconducting (SC) phase fluctuations,%
\cite{Emery:1995} 
strong magnetic fluctuations near the antiferromagnetic (AF)
instability,\cite{%
Schmalian:1998} and so on.
At present there is no consensus on the origin of the pseudogap and 
its relationship with the SC long-range order.
Among several scenarios, the 
slave boson (or fermion) mean-field theory%
\cite{Anderson:1987,Baskaran:1987}
may shed some insight into the 
problem.  
This is because the pseudogap is closely related to a spin gap, 
the predicted phase diagram is, at least, qualitatively 
consistent with experiments, and furthermore
it starts from the microscopic model ($t-J$ Hamiltonian) 
as opposed to other
phenomenological models.
On the other hand, 
the recombination of a spinon and a holon into a physical electron is 
nontrivial, and also the constraint of no-double-occupancy at each site
is difficult to impose at a microscopic level.
In deriving a local spin-singlet amplitude which may be 
responsible for the pseudogap behavior,
in this paper we use 
the Hubbard Hamiltonian 
instead of the $t-J$ Hamiltonian.
Local spin-singlet amplitude is induced directly  
from short-range spin correlations in the normal state 
and its strength is quantified 
near half-filling.
In this paper short-range spin correlation means that when site $i$ is
occupied by an electron with up-spin (or down-spin),
then the nearest sites predominantly by electrons with the {\em opposite} spin.

   We start by defining the one-band Hubbard model
proposed by  
Anderson\cite{Anderson:1987}
as the simplest model which might capture the correct 
low energy physics of copper oxides. 
The Hubbard model is described by the Hamiltonian 
where $c_{i,\sigma}$ destroys an electron at site $i$ with spin $\sigma$
on a two-dimensional square lattice 
\begin{eqnarray}
 H = -t\sum_{<i,j>,\sigma}c^{\dag}_{i,\sigma}
      c_{j,\sigma}
    +U\sum_{i}
     c^{\dag}_{i,\uparrow}
         c_{i,\uparrow}
     c^{\dag}_{i,\downarrow}
         c_{i,\downarrow} \; .
                                                           \label{eq10}
\end{eqnarray}
$t$ is a hopping parameter between nearest neighbors $<i,j>$ and 
$U$ denotes local Coulomb repulsion.
It is believed that the realistic strength of the Coulomb repulsion
lies in between the weak and strong coupling regimes, namely,
$U \sim W-2W$ where $W$ is the bandwidth of $8t$ in two dimensions.

   As a first step to the microscopic understanding of the pseudogap,
it is important to answer a more fundamental question:
How can spin-singlet tendency\cite{Comment5} 
appear directly from the Coulomb repulsion,
presumably without the exchange of bosonic degrees of freedom
such as spin fluctuations\cite{Bickers:1989}?
As a simple example to illustrate this point qualitatively,
let us consider the $U \gg t$ limit at half-filling where 
only an up-spin or down-spin electron is allowed at each site.
The typical spin-singlet structure (with $d$-wave form factor) 
in which we are interested in this paper is 
$\Delta_{g}(i)  =  \frac{1}{2}\sum_{\delta}g(\delta)
       (c_{i+\delta,\uparrow}        c_{i,\downarrow}
       -c_{i+\delta,\downarrow}      c_{i,\uparrow})$, 
where $g(\delta)$ is an appropriate 
structure factor.
For a d-wave symmetry, for instance,
\begin{eqnarray}
 g(\delta) = \left\{ \begin{array}{lll}
                          1/2     &  \mbox{if $\delta=(\pm 1,0)$} , \\
                         -1/2     &  \mbox{if $\delta=(0,\pm 1)$} , \\
                           0      &  \mbox{if otherwise} .
                       \end{array}
               \right.
                                                        \label{eq20}
\end{eqnarray}
Then 
the local spin-singlet amplitude
for strongly correlated electrons
($\langle |\Delta_{g}(i)| \rangle =(1 \times 1 + 0 \times 0)/2 = 1/2$)  
is increased 
over its noninteracting value  
($\langle |\Delta_{g}(i)| \rangle_{0}= (1/2 \times 1/2 + 1/2 \times 1/2)/2 
= 1/4$ 
for both spins).

   To make this argument more quantitative,  
we introduce a spin-singlet correlation function\cite{Comment7}
\begin{eqnarray}
\chi_{g}(i,\tau) = \langle T_{\tau} \Delta_{g}(i,\tau)
                                    \Delta^{\dag}_{g}(0,0) \rangle \; ,
                                                        \label{eq25}
\end{eqnarray}
where $T_{\tau}$ is the imaginary time ordering operator.
In this paper we consider only the {\em local} spin-singlet amplitude
$\langle |\Delta_{g}(0)|^{2} \rangle$, which may be obtained by
$\chi_{g}(i \rightarrow 0, \tau \rightarrow 0^{-})$.
Now we examine whether there is an increase in 
$\langle |\Delta_{g}(0)|^{2} \rangle$
for strongly correlated electrons with respect to 
$\langle |\Delta_{g}(0)|^{2} \rangle_{0}$
for the noninteracting electrons,  
which is similar in spirit to 
a renormalization group (RG) approach\cite{Schulz:1987}. 
As already illustrated in the previous paragraph, the increase in 
$\langle |\Delta_{g}(0)|^{2} \rangle$
for strongly correlated electrons 
crucially depends on the short-range spin correlations. 
Although there exists no controlled way of obtaining the wave function 
in general,  
certain local or short-range {\em static} quantities such as
double occupancy (or equivalently local spin amplitude) 
or the nearest neighbor correlations
are reasonably
well captured by the mean-field state with AF order.
Before going further, it is important to establish the validity of
the current approximation by explicitly comparing 
the above quantities calculated in the mean-field state (with AF order)
with available quantum Monte Carlo (QMC) results.

\begin{figure}
 \vbox to 7.0cm {\vss\hbox to -5.0cm
 {\hss\
       {\includegraphics{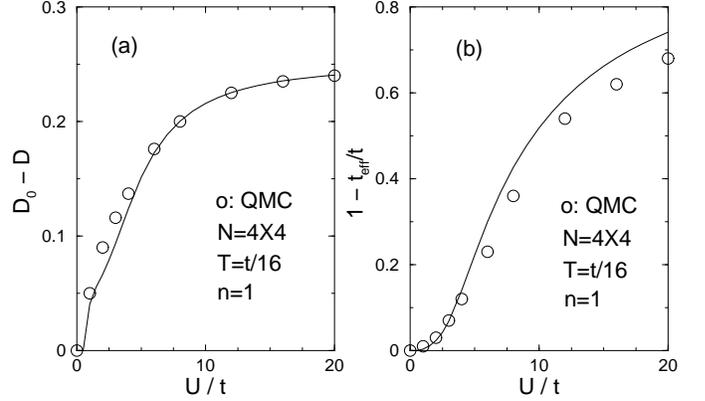}
       }
  \hss}
 }
\caption{(a) Double occupancy
$D=\langle n_{i,\uparrow} n_{i,\downarrow} \rangle$ and
 (b) nearest neighbor correlation
$t_{eff}/t=\langle c^{\dag}_{i,\sigma}c_{j,\sigma} \rangle_{U} /
           \langle c^{\dag}_{i,\sigma}c_{j,\sigma} \rangle_{U=0}$
calculated in the
mean-field state with AF order (solid curve) and in the QMC simulations
by White {\it et al.}\protect\cite{White:1989} (open circles)
for $n=1$, $N=4 \times 4$, and $T=t/16$.
For purely interaction induced effect, $D_{0}$ and unity are subtracted
from $D$ and $t_{eff}/t$, respectively.}
\label{fig1.5}
\end{figure}

   The double occupancy
$D=\langle n_{i,\uparrow} n_{i,\downarrow} \rangle$ plays an important
role in gauging the degree of strong correlations in the Hubbard-type
model. For instance, $D \rightarrow 0$ at half-filling for
$U \rightarrow \infty$, while $D_{0}$ evaluated in the noninteracting
state is $(n/2)^{2}=0.25$.
In Fig.~\ref{fig1.5}(a) our calculations (solid curve) are compared
with (virtually exact) QMC results (open circles)
by White {\it et al.}\cite{White:1989}
for $n=1$, $N=4 \times 4$, and $T=t/16$.
For purely interaction induced effect, $D_{0}$ is subtracted
from $D$. Above $U \simeq 3t$ the agreement with QMC data
is increasingly better and for $U \ge 8t$ the two results are
almost indistinguishable.
For the nearest neighbor correlations between $i$ and $j$,
we calculate
$t_{eff}/t=\langle c^{\dag}_{i,\sigma}c_{j,\sigma} \rangle_{U} /
           \langle c^{\dag}_{i,\sigma}c_{j,\sigma} \rangle_{U=0}$
for comparison with available QMC results.
Again $t_{eff}/t$ evaluated in the noninteracting state, namely,
unity is subtracted from $t_{eff}/t$ for purely interaction induced effect.
Figure~\ref{fig1.5}(b) shows our results evaluated in the mean-field
state (solid curve) together with
QMC results (open circles)
by White {\it et al.}\cite{White:1989}
From intermediate to very strong coupling, the deviation with QMC results is
less than $10 \%$ level.
In fact the interaction induced local spin-singlet amplitude
 $\langle |\Delta_{g}(0)|^{2} \rangle$
-$\langle |\Delta_{g}(0)|^{2} \rangle_{0}$
involves mainly nearest neighbor static correlations between
electrons.
Note that the spin density wave (SDW) approximation will be used below
{\em only to capture the reasonable local and short-range correlations}  
between electrons.

   We should also point out some limitations in this approach.
The physical reason that the nearest neighbor correlation
is overestimated (by up to $10 \%$) compared with QMC data is as follows.
Suppose site $i$ is
occupied by an electron with up-spin in the strong coupling limit.
Then the nearest sites are occupied by down-spin electrons as majority
and also by up-spin electrons as minority. As the distance from
site $i$ increases, the effective polarization of electron spins
decreases in magnitude.
In the mean-field state with AF order, however,
the effective polarization is the same for all the distances from site $i$.
This means that physical quantities evaluated in the mean-field state
become progressively overestimated for increasingly distant correlations.
Another limitation is that the mean-field critical
doping $x_{c}$ where AF mean-field order vanishes 
increases with increasing $U$.
This is inconsistent with more advanced treatment of the Hubbard
model\cite{Kotliar:1986} where with increasing $U$,
$x_{c}$ decreases after reaching its maximum around at $x \sim 0.21$.
Hence our approximation based on the Hubbard model is valid near half-filling
(probably up to $x \sim 0.1-0.15$).
For the $t-J$ Hamiltonian
the current approximation will be valid even far away
from half-filling,
because this strong coupling feature is already taken into account there.

\begin{figure}
 \vbox to 7.0cm {\vss\hbox to -5.0cm
 {\hss\
       {\includegraphics{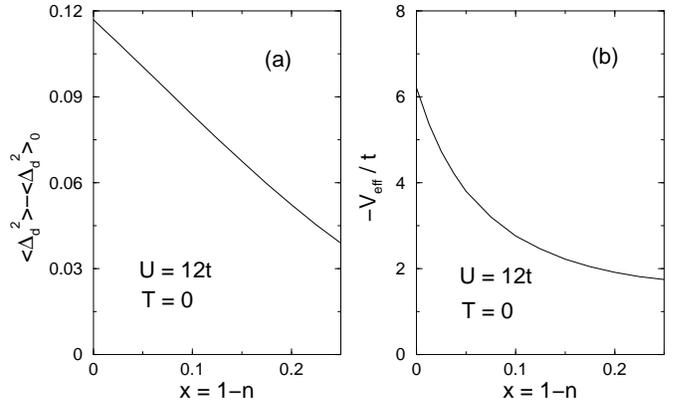}
       }
  \hss}
 }
\caption{
(a) Local spin-singlet amplitude $\langle |\Delta_{d}|^{2} \rangle$
    calculated in the mean-field
    state with AF order for $U=12t$, $T=0$.
    $\langle |\Delta_{d}|^{2} \rangle_{0}$ is subtracted for
    purely interaction induced effect.
(b) Effective strength $V_{ind}$ of the local spin-singlet amplitude
    induced by short-range
    spin correlations for $U=12t$, $T=0$.}
\label{fig1.8}
\end{figure}

   Then for $d$-wave type symmetries, 
the local spin-singlet amplitude  
$\langle |\Delta_{g}(0)|^{2} \rangle$ 
evaluated in the mean-state with AF order 
becomes 
\begin{eqnarray}
& &  \frac{n}{4N}\sum^{'}_{\vec{k}}\phi_{g}^{2}(\vec{k})
     [f(E_{-}(\vec{k}))
     +f(E_{+}(\vec{k}))]
-sign[\frac{\phi_{g} (\vec{k}+\vec{Q})}{\phi_{g}(\vec{k})}]
                                             \nonumber  \\
&\times& \frac{\Delta_{sdw}}{2UN}\sum^{'}_{\vec{k}}\phi_{g}^{2}(\vec{k})
     \frac{\Delta_{sdw}}
          {\lambda(\vec{k})}
     [f(E_{-}(\vec{k}))
     -f(E_{+}(\vec{k}))]  
                           \; ,
                                                           \label{eq30}
\end{eqnarray}
where 
\begin{eqnarray}
\lambda(\vec{k}) & = & {\sqrt{((\varepsilon(\vec{k})
                -\varepsilon(\vec{k}+\vec{Q}))/2)^{2}
                +\Delta_{sdw}^{2}}}  \; ,
                                             \nonumber  \\
E_{\pm}(\vec{k}) & = &
  (\varepsilon(\vec{k})+\varepsilon(\vec{k}+\vec{Q}))/2
   \pm \lambda(\vec{k})
                           \; .
                                                           \label{eq35}
\end{eqnarray}
$\varepsilon(\vec{k})=-2t(\cos k_{x}+\cos k_{y})-\mu$ for
nearest neighbor hopping, 
$\mu$ is the chemical potential controlling
the particle density $n$,
$N$ the total number of lattice sites,
$f(E)$ the Fermi-Dirac distribution function,
$\vec{Q}$ the AF wave vector $(\pi,\pi)$,
$\phi_{g}(\vec{k})$ the Fourier transform of $g_{\delta}(i)$, and 
the summation accompanied by the prime symbol is over wave vectors
in half of the first Brillouin zone.
The SDW gap $\Delta_{sdw}$ and the chemical potential $\mu$ are 
self-consistently determined through the gap and number equations
for given $U$, $T$ and $n$.
Figure~\ref{fig1.8}(a) shows the local spin-singlet amplitude
$\langle |\Delta_{d}|^{2} \rangle$ 
subtracted by that evaluated 
in the noninteracting state for purely interaction induced effect.
The most interesting finding is that the local spin-singlet amplitude
is induced directly by short-range spin correlations without any 
explicit driving (attractive) interaction.
The short-range spin correlations are ultimately ascribed to the 
strong local Coulomb repulsion $U$.
The interaction induced $\langle |\Delta_{d}|^{2} \rangle$
rapidly decreases with increasing doping, and if the second limitation
in the current approximation mentioned above is properly corrected, 
it will decrease even faster with doping.

   Because the value of the first term is nearly the same as 
$\langle |\Delta^{+}_{g}(0)|^{2} \rangle_{0}$
for the noninteracting electrons,
the positivity of the second term\cite{Comment10} 
leads to an increase in the local spin-singlet amplitude.
For $d_{xy}$
($\phi(\vec{k})=2\sin k_{x} \sin k_{y}$) symmetry, for example,
the local spin-singlet amplitude is even suppressed due to 
$\phi (\vec{k}+\vec{Q})=\phi (\vec{k})$.
We found that 
among several symmetries including local s-wave, extended s-wave,
$d_{xy}$, and 
$d_{x^2-y^2}$,
the $d_{x^2-y^2}$ ($\phi_{d} (\vec{k})=\cos k_{x} - \cos k_{y}$) symmetry
shows the largest increase in the local spin-singlet amplitude.
Likewise the local spin amplitude $\langle (S_{z}(i))^{2} \rangle$
is also induced from the same 
Coulomb repulsion.

   In order to determine the effective strength of
the induced local spin-singlet amplitude (with $d$-wave symmetry)
corresponding to Fig.~\ref{fig1.8}(a),
we consider a Hamiltonian 
\begin{eqnarray}
 H = -t\sum_{<i,j>,\sigma}c^{+}_{i,\sigma}
      c_{j,\sigma}
     + V_{ind}\sum_{i}
          \Delta^{+}_{d}(i) \Delta_{d}(i)
                                                           \label{eq37}
\end{eqnarray}
with the same parameters ($t$, $T$, and $n$) as before.
Now the strategy is to find $V_{ind}$ in such a manner that 
$V_{ind}$ gives the same interaction induced local spin-singlet amplitude 
 $\langle |\Delta_{d}|^{2} \rangle$ 
-$\langle |\Delta_{d}|^{2} \rangle_{0}$ 
as in Fig.~\ref{fig1.8}(a).
Please note that $V_{ind}$ is the effective strength of
the induced local spin-singlet amplitude, but not attractive
interaction yielding a local SC pair 
in the Bose-Einstein limit.
As before,
we may gauge the validity of
a given approximation (BCS approximation in this case) by explicitly comparing 
some local and short-range static quantities
calculated in that approximation 
with available numerical data for the above 
Hamiltonian.
Unfortunately there exist no such numerical results. 
Then one may use general knowledge which has been obtained 
through study on the repulsive and attractive Hubbard models
in the SDW and BCS approximations, respectively.
The SDW approximation for the repulsive Hubbard model which was 
discussed in details in the previous paragraphs, suggests that   
for intermediate to strong coupling
the mean-field approximation reasonably well captures
both local and short-range static quantities.
For the BCS approximation of the attractive Hubbard 
model\cite{Kyung:2000-6}, it has also been shown that the local static 
quantity (double occupancy) in that approximation is in good agreement with 
QMC data for $|U| \ge 3t$.
However, the numerical results of nearest neighbor correlations 
in the attractive Hubbard model are not available in the literature
to our knowledge.
In spite of lack of complete numerical data
for the attractive Hubbard model and for the Hamiltonian Eq.~\ref{eq37},
one may {\em safely} expect that the BCS mean-field approximation 
to Eq.~\ref{eq37} will give at least reasonable local and short-range static  
quantities for intermediate to strong coupling.
Note that the mean-field (BCS) approximation will be used below
{\em only to capture the reasonable local and short-range correlations}  
between electrons, as before in the SDW approximation.

   In the BCS approximation for the above Hamiltonian,
$\langle |\Delta_{d}(0)|^{2} \rangle$ is given as 
\begin{eqnarray}
& &  \frac{n}{4N}\sum_{\vec{k}}\phi_{d}^{2}(\vec{k})
     [\frac{1}{2}-\frac{\varepsilon(\vec{k})}
     {2E(\vec{k})}\tanh (E(\vec{k})/2T)]
                                             \nonumber  \\
&+& [\frac{1}{N}\sum_{\vec{k}}\phi_{d}(\vec{k})
     \frac{\Delta_{d}(\vec{k})}
               {2E(\vec{k})}\tanh (E(\vec{k})/2T)]^{2}
                           \; ,
                                                           \label{eq40}
\end{eqnarray}
where 
$E(\vec{k})=\sqrt{\varepsilon^{2}(\vec{k})
                +\Delta_{d}^{2} (\vec{k})}$.
As before, the d-wave gap $\Delta_{d}(\vec{k})=\Delta \phi_{d}(\vec{k})$ and  
the chemical potential $\mu$ are 
self-consistently determined through the gap and number equations
for given $V_{ind}$, $T$ and $n$.
In Fig.~\ref{fig1.8}(b) $V_{ind}$ 
is plotted 
as a function of doping $x=1-n$ for $T=0$ and $U=1.5W=12t$ near half-filling.
The BCS approximation ($|V_{ind}| \ge 3t$) is expected to be accurate 
for $x \le 0.1$ and to be less accurate beyond it. 
The strength of the induced local spin-singlet amplitude 
rapidly decreases with doping,
just as the induced AF correlations do.
Beyond $x \sim 0.1 - 0.15$,
$|V_{ind}|$ will decrease even faster when the strong 
coupling effect is correctly included, as was mentioned 
in Fig.~\ref{fig1.8}(a).
For $U \gg t$ at $n=1$, $V_{ind}$ saturates to be 
$-7.64t$ and 
$\langle |\Delta_{d}(0)|^{2} \rangle \rightarrow 0.25$,  
consistent with our previous result 
($\langle |\Delta_{d}(0)| \rangle = 0.5$) based on qualitative 
argument.
$V_{ind}$ and its doping dependence
have some interesting consequences which may answer some of
the puzzling issues in the high temperature superconductors.

   As a first application of the present approach, we attack a
problem of the energy dispersion and its d-wave modulation 
in the insulating cuprates, 
Sr$_{2}$CuO$_{2}$Cl$_{2}$ and 
Ca$_{2}$CuO$_{2}$Cl$_{2}$.
Recent ARPES experiments for an insulating cuprate
Sr$_{2}$CuO$_{2}$Cl$_{2}$\cite{Wells:1995}
clearly show that
the near isotropy and the overall band dispersion 
along $(\pi/2,\pi/2)-(\pi,0)$ and
$(\pi/2,\pi/2)-(0,0)$ cannot be explained by considering 
only AF order or its fluctuations, 
unless some adjustable fitting parameters such as $t'$ and $t''$ are
introduced.
Furthermore 
a d-wave-like modulation of the insulating gap 
in Ca$_{2}$CuO$_{2}$Cl$_{2}$\cite{Ronning:1998}
is totally mysterious from that point of view.
Because the induced local spin-singlet and 
spin amplitudes increase with decreasing doping,
near half-filling and at low temperatures,
both short-range spin-singlet and AF fluctuations are strong
and coexist, leading to  
a resonating valence
bond (RVB) type state\cite{Anderson:1987}. 
Our general experience\cite{Vilk:1997}  
tells that as long as the energy dispersion 
is concerned, it is basically identical 
in a strongly fluctuating state (or in the pseudogap state) and 
in a long-range ordered state. 
The only difference is that in the former state
the gap is filled with some spectral weight (pseudogap)
and the spectral function is spilt into two (relatively broad) peaks instead of 
two delta functions.
Thus at half-filling and at low temperatures,
a hole strongly interacts with both short-range spin-singlet and AF 
fluctuations, yielding 
an energy dispersion similar to that  
in the coexistence state of the SC and AF 
long-range order.\cite{Kyung:2000-2}
The energy dispersion is then given by 
\begin{eqnarray}
  \sqrt{\varepsilon^{2}(\vec{k}) + \Delta^{2}_{sdw}
        + \Delta^{2}(\cos k_{x} -\cos k_{y})^{2}}
                                            \; .
                                                           \label{eq50}
\end{eqnarray}

\begin{figure}
 \vbox to 7.0cm {\vss\hbox to -5.0cm
 {\hss\
       {\includegraphics{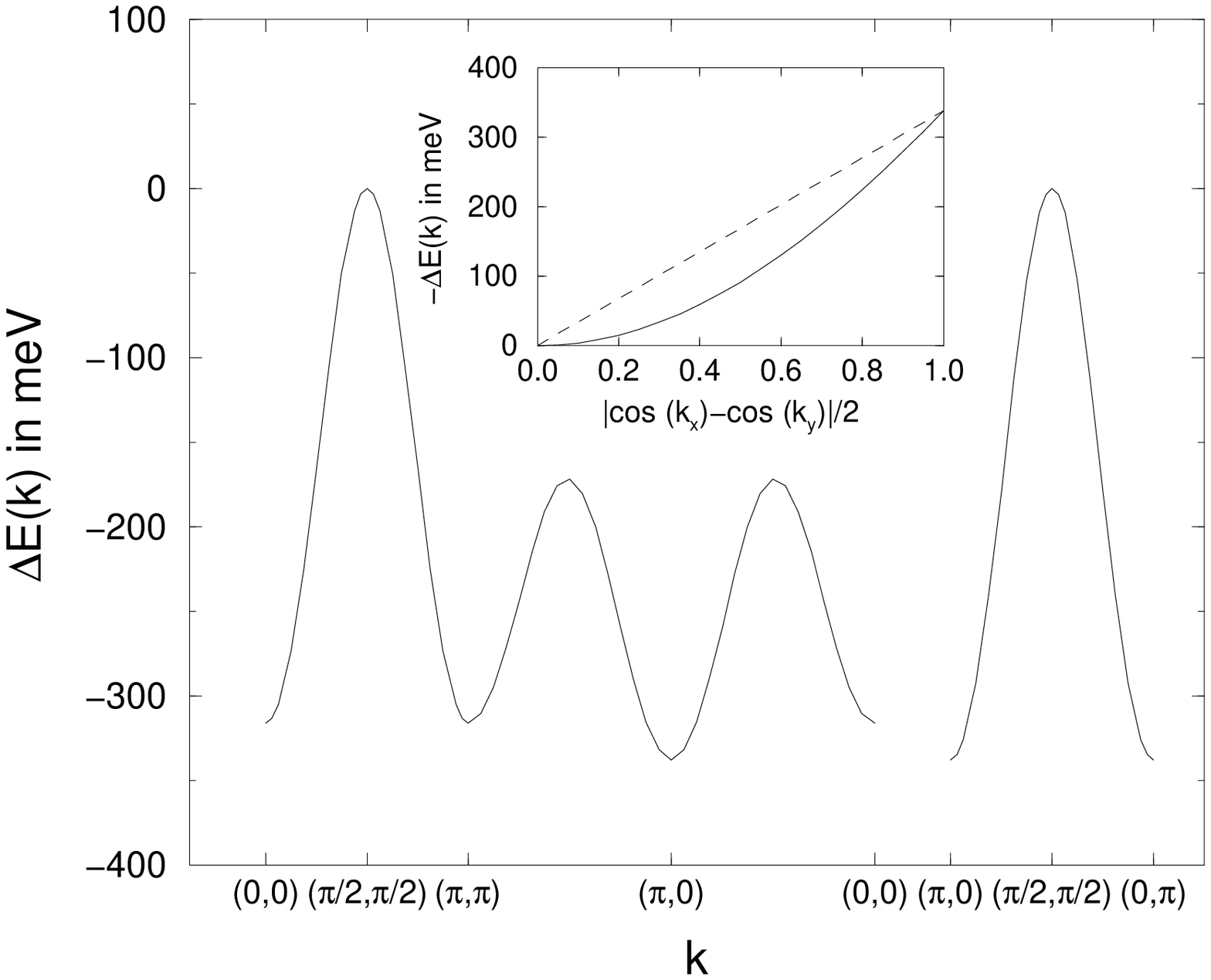}
       }
  \hss}
 }
\caption{Calculated energy dispersion for an insulating cuprate
         Ca$_{2}$CuO$_{2}$Cl$_{2}$. The dispersion is measured  
         from $(\pi/2,\pi/2)$ point with $\Delta_{sdw} = U/2$
         and $\Delta=2.07t$. The solid and dashed curves in the inset 
         denote the dispersions calculated from our approach, and 
         from the flux phase in the $t-J$ model, respectively.}
\label{fig2}
\end{figure}

   From the overall bandwidth ($\simeq$ 320 meV) experimentally measured 
from $(\pi/2,\pi/2)$ to $(0,0)$ 
along which $\phi_{d}(\vec{k})$ vanishes, the Coulomb
repulsion $U$ (and $\Delta_{sdw} \simeq U/2$) is determined 
to be $11.4t$ for $t=$250 meV.
For $U=11.4t$, $n=1$ and $T=0$, the effective strength $V_{ind}$ of  
the induced spin-singlet amplitude 
is completely determined to give  
$V_{ind}=-6.09t$ and $\Delta=2.07t$.
Figure~\ref{fig2} clearly demonstrates that 
the energy dispersion and its near isotropy
along $(\pi/2,\pi/2)-(\pi,0)$ and
$(\pi/2,\pi/2)-(0,0)$ are in excellent agreement with experiments
{\em without any adjustable parameters}.
Our energy dispersion 
along $(\pi/2,\pi/2)-(\pi,0)$ is proportional to 
$(\cos k_{x} -\cos k_{y})^{2}$ (solid curve in the inset) at low energies 
as opposed to
$|\cos k_{x} -\cos k_{y}|$ (dashed curve) predicted from the 
flux-phase in the $t-J$ model.\cite{Kotliar:1988}
Recent experiments show a significant deviation of the energy dispersion from 
the cusp-like form
along $(\pi/2,\pi/2)-(\pi,0)$.
A similar result to ours was recently obtained in the context of SO(5) 
symmetry.\cite{Zacher:2000} 
The state with
a spin gap (or pseudogap) of $(\Delta \phi_{d})^{2}/U \simeq J\phi_{d}^{2}$ 
and also with a Mott-Hubbard gap of order $U$ at half-filling
is expected to continuously evolve into a state with a relatively smaller
spin gap (or pseudogap) but without a charge gap away from half-filling.

   In this paper we have considered only the local spin-singlet
amplitude induced by short-range spin correlations
and its consequences.
The long-range $d$-wave superconductivity is 
beyond the scope of the present approach.
How local spin-singlets acquire local SC phases and 
eventually establish their long-range phase
coherence is a challenging problem
to the theory of high temperature superconductivity.
Quantitative analytical study of the Hubbard model 
in the physically relevant regime
and of the interplay between AF and $d$-wave
pairing correlations beyond a mean-field level is not available
at present, mainly due to the absence of a small parameter.
Certainly it is a subject for future study.
It is also desirable to perform numerical simulations for
a Hamiltonian with $d$-wave symmetry such as Eq.~\ref{eq37}
if it is possible.
Then the mean-field BCS approximation used here to obtain
reasonable local and short-range static correlations
between electrons will be tested for its range of validity.

   In summary, from the microscopic Hubbard Hamiltonian,
the local spin-singlet amplitude due to short-range spin correlations
was explicitly extracted
and its strength was quantified near half-filling.
As a first application of the present approach,
a problem of the energy dispersion and its d-wave modulation
in the insulating cuprates,
Sr$_{2}$CuO$_{2}$Cl$_{2}$ and
Ca$_{2}$CuO$_{2}$Cl$_{2}$
was studied.
Without any adjustable parameters,
most puzzling issues on the energy dispersion were
naturally and quantitatively explained within the present approach.

   The author would like to thank A. M. Tremblay for numerous 
help and discussions throughout this work. 
The author also thanks C. Bourbonnais, E. Dagotto, R. Gooding, M. Norman, 
D. S\'{e}n\'{e}chal, and W. Stephan
for stimulating discussions.
The present work was supported by a grant from the Natural Sciences and
Engineering Research Council (NSERC) of Canada and the Fonds pour la
formation de Chercheurs et d'Aide \`a la Recherche (FCAR) of the Qu\'ebec
government.
%
%
%

%
%
%
\newpage

\begin{references}
\bibitem{Comment5} 
Spin-singlet tendency is not necessarily associated with superconductivity.   
As will be demonstrated shortly, local spin-singlet can arise entirely from 
short-range spin correlations without any explicit attractive interaction.
\bibitem{Comment7}
The spin-singlet correlation function (with
$d$-wave symmetry) is more general than
the $d$-wave pair correlation function
in the sense that
the spin-singlet correlations can be also induced by
short-range spin correlations even in the absence of
pairing interactions.
\bibitem{Comment10} 
For local $s$-wave and extended $s$-wave symmetries, there are additional
contributions to the above equation. But the second term 
is still the major factor determining
whether 
$\langle |\Delta_{g}(0)|^{2} \rangle 
 \langle |\Delta_{g}(0)|^{2} \rangle_{0}$ or not.
\end{references}
\end{document}